\documentclass{llncs}

\usepackage{graphicx}
\usepackage{ctable}
\usepackage{amssymb, amsmath}
\usepackage{bm}
\usepackage{siunitx}
\usepackage{physics}

\usepackage{algorithm}
\usepackage[noend]{algpseudocode}

\usepackage{graphicx}

\usepackage{cite}
\begin{document}

\title{Acceptance Rates of Invertible Neural Networks on Electron Spectra from Near-Critical Laser-Plasmas: A Comparison}

\author{Thomas Miethlinger\inst{1,2} \and
Nico Hoffmann\inst{1} \and
Thomas Kluge\inst{1}}

\authorrunning{T. Miethlinger et al.}

\institute{Helmholtz-Zentrum Dresden-Rossendorf, 01328 Dresden, Germany \email{\{t.miethlinger,n.hoffmann,t.kluge\}@hzdr.de} \and
Technische Universität Dresden, 01069 Dresden, Germany}

\maketitle

\begin{abstract}
While the interaction of ultra-intense ultra-short laser pulses with near- and overcritical plasmas cannot be directly observed, experimentally accessible quantities (observables) often only indirectly give information about the underlying plasma dynamics.
Furthermore, the information provided by observables is incomplete, making the inverse problem highly ambiguous.
Therefore, in order to infer plasma dynamics as well as experimental parameter, the full distribution over parameters given an observation needs to considered, requiring that models are flexible and account for the information lost in the forward process.
Invertible Neural Networks (INNs) have been designed to efficiently model both the forward and inverse process, providing the full conditional posterior given a specific measurement.
In this work, we benchmark INNs and standard statistical methods on synthetic electron spectra.
First, we provide experimental results with respect to the acceptance rate, where our results show increases in acceptance rates up to a factor of 10.
Additionally, we show that this increased acceptance rate also results in an increased speed-up for INNs to the same extent.
Lastly, we propose a composite algorithm that utilizes INNs and promises low runtimes while preserving high accuracy.

\end{abstract}

\setcounter{footnote}{0}
\section{Introduction}
Relativistic plasmas driven by ultra-intense ultra-short laser pulses are currently increasingly investigated due to various prospective applications in e.g. medicine, materials science and laboratory astrophysics.
While the dynamics of \textit{underdense} plasmas, i.e. plasmas with electron density $n_\mathrm{e}$ smaller than the critical plasma density $n_\mathrm{c}$, can in principle be studied with optical methods as incoming light there is mostly transmitted, the situation is much more difficult for \textit{near-critical} (mostly absorption, $n_\mathrm{e} \approx n_\mathrm{c}$) and \textit{overdense} (mostly reflection, $n_\mathrm{e} > n_\mathrm{c}$) plasmas.
Indeed, inferring experimental parameter values and consequently determining the relevant underlying plasma dynamics is highly elaborate, heavily depending on comparisons with observables computed from plasma simulations where typically the Particle-in-Cell (PIC) method is employed~\cite{birdsall2018plasma,hockney2021computer}. 
Furthermore, the information provided by observables is incomplete in the sense that multiple experimental parameter and plasma dynamics can cause the same values for observables,
but also retrieving information is regarded non-trivial since this process usually depends on fitting (scalar) quantities to analytical expressions which have been derived under strong assumptions.

This is, depending on the context, also the case for the \textit{electron spectrum}, which counts the number of electrons $\dd{N}_\mathrm{e}$ in an energy interval $\dd{E}$:
\begin{equation}
    f_\mathrm{e}(E) := \frac{\dd{N}_\mathrm{e}}{\dd{E}}.
\end{equation}
In this work, we study \textit{laser-driven ion acceleration}~\cite{wilks2001energetic,macchi2017review}.
In this research field, where objectives concern ion-related properties, the electron spectrum is a secondary quantity that is sometimes measured in conjunction with the ion spectrum.
Being a high-dimensional vector, however, the electron spectrum is difficult to interpret, infer parameters and draw conclusions from.
Typically, one resorts to computing the \textit{mean (kinetic) energy} of the laser-driven electrons\footnote{Sometimes (unfortunately) also called \textit{electron temperature}.}:
\begin{equation}
    T_\mathrm{e} := \frac{\int_{E_\mathrm{e}^\mathrm{laser}}^\infty{E f_\mathrm{e}(E) \dd{E}}}{\int_{E_\mathrm{e}^\mathrm{laser}}^\infty{f_\mathrm{e}(E) \dd{E}}},
\end{equation}
where the lower integration boundary $E_\mathrm{e}^\mathrm{laser}$ is introduced to distinguish between electrons in thermal equilibrium and laser-driven electrons exhibiting an exponential distribution for high energies.
One can then show, using analytical considerations, that the \textit{ion cutoff energy}\footnote{Or, equivalently, the maximum ion energy in a laser-driven ion spectrum.} scales linearly with the mean kinetic energy of the electrons, $E_\mathrm{i}^\mathrm{max} \propto T_\mathrm{e}$, and that the mean kinetic energy of the electrons itself mostly depends on the laser intensity $I$: $T_\mathrm{e}=T_\mathrm{e}(I)$~\cite{mora2003plasma,kluge2011electron}.

However, retrieving information from the electron spectrum beyond $T_\mathrm{e}$ in general requires an automatized, data-driven approach.
This is all the more the case because PIC simulations are computationally (potentially very) expensive, which motivates researchers to this day to improve PIC simulation codes, for example, algorithmically or by improved hardware utilization~\cite{burau2010picongpu,derouillat2018}.
Therefore, employing machine learning (ML) algorithms and ML-based surrogate models is essential to decrease the overall computational effort which would otherwise be needed due to the necessity of performing an excessive amount of simulations.
For example, Djordjević et al. used deep learning to predict the time evolution of ion cutoff energies and electron mean kinetic energies in overdense laser-ion acceleration~\cite{djordjevic2021modeling}.

\section{PIC Simulation Setup and Data Generation}\label{sec:sim_data}
In this work, we employ the PIC-code \textit{Smilei}~\cite{derouillat2018} to generate data for our ML models.
Since the predominant acceleration mechanism in laser-ion acceleration is target-normal sheath-acceleration (TNSA), which is a one-dimensional physical effect, and in order to significantly reduce the computational effort for this study, we use a narrow simulation box with $240\si{\um} \times 0.2\si{\um}$ and impose periodic boundary conditions in the $y$-direction~\cite{wilks2001energetic}.
The cell size is $\Delta x = \Delta y = \frac{\lambda_\mathrm{0}}{\mathrm{resolution}} = \frac{800 \si{nm}}{64} = 12.5 \si{nm}$, and the time step is $\Delta t = 0.995 \frac{1}{\sqrt{2}} \frac{\Delta x}{c}$ corresponding to a Courant–Friedrichs–Lewy (CFL) value of $0.995$.
We initialize our plasma with 50 particles per cell.
The target is a pre-expanded hydrogen foil with thickness $D$.
Pre-plasma with exponential scale length $\ell$ is included at the front side of the target such that the density reaches a maximum $n_0$ at $x_\mathrm{f}=100\si{\um}$.
Moreover, the pre-plasma is cut-off where the density is less than $0.01 n_\mathrm{c}$, i.e. $\forall x: n(x) < 0.01 n_\mathrm{c} \implies n(x) = 0$.
The back side is not pre-expanded, i.e. the density is step-function-like shaped.
The laser pulse is a Gaussian with full width at half maximum (FWHM) $\tau$ and normalized vector potential $a_0 = E/E_0 = E/(e^{-1} m_\mathrm{e} c \omega_0)$, where $E$ is the corresponding electric field, $e$ is the elementary charge, $m_\mathrm{e}$ the electron mass, $c$ the speed of light and $\omega_0 = 2\pi \frac{c}{\lambda_0}$ the angular frequency corresponding to the laser's central wavelength $\lambda_0 = 800 \si{nm}$.

In this work, we performed 5000 simulations in total, varying the five parameters $a_0, \tau, n_0, D$ and $\ell$.
An overview of the parameter space that we studied in this work is given in Table~\ref{tab:sim_params}.
Thus, the laser intensity is in the range between $10^{20} \leq I(a_0)/\mathrm{W}\si{\cm}^{-2} \leq 10^{21}$.
\begin{table}[h]
\centering
\caption{Parameter space for PIC simulations.}
\begin{tabular}{c|c|c|c|c|c} 
\specialrule{.1em}{.05em}{.05em} 
\textbf{Quantity} & \textbf{Symbol} & \textbf{Unit} & \textbf{Min} & \textbf{Max} & \textbf{Scaling} \\
\specialrule{.1em}{.05em}{.05em}
Normalized vector potential & $a_0$ & 1 & 6.8 & 21.5 & linear \\
\specialrule{.05em}{.05em}{.05em}
Full width at half maximum & $\tau$ & \si{\fs} & 25 & 50 & linear \\
\specialrule{.05em}{.05em}{.05em}
Number density (bulk) & $n_0$ & $n_\mathrm{c}$ & 15 & 60 & linear \\
\specialrule{.05em}{.05em}{.05em}
Target thickness & $D$ & \si{\um} & 0.25 & 5 & linear \\
\specialrule{.05em}{.05em}{.05em}
Pre-plasma scale length & $\ell$ & \si{\um} & 0.01 & 1 & square \\
\specialrule{.1em}{.05em}{.05em}
\end{tabular}
\label{tab:sim_params}
\end{table}\newline
Since ML usually strongly benefits from using normalized and/or standardized values, we designed our experiments as follows:
\begin{itemize}
    \item In an effort to have our simulations as space-filling as possible in parameter space, we obtained our parameter vectors $\vb{x}=\left[x_1, ..., x_5 \right]^\mathsf{T}$ from a low-discrepency sequence.
    In particular, we used the \textit{Halton sequence}, which is a common low-discrepancy sequence used in Monte Carlo integration and design of experiment, with dimension $n_{\vb{x}}=5$ and support $\hat{x}_k\in \left[0, 1\right] \forall k \in \{1,...,5\}$~\cite{niederreiter1992random}.
    \item Since $\ell$ spans two orders of magnitude, we account for that by using a nonlinear transformation to obtain the parameter values $x_k$ as used in our simulations.
    This can be expressed as follows:
    \begin{equation}
        x_k = \hat{x}_k^{\mathrm{s}} (x_k^\mathrm{max} - x_k^\mathrm{min}) + x_k^\mathrm{min},
    \end{equation}
    where $x_k^\mathrm{min}$ and $x_k^\mathrm{max}$ refer to the parameter range of the $k^\mathrm{th}$ parameter $x_k$, as defined in Table~\ref{tab:sim_params}, and where the scaling exponent $\mathrm{s}$ refers to $\mathrm{s}=1$ for $k\leq 4$ (linear), and $\mathrm{s}=2$ for $k=5$ (squared scaling for $\ell$).
\end{itemize}

\begin{figure}[t]
\includegraphics[width=\textwidth]{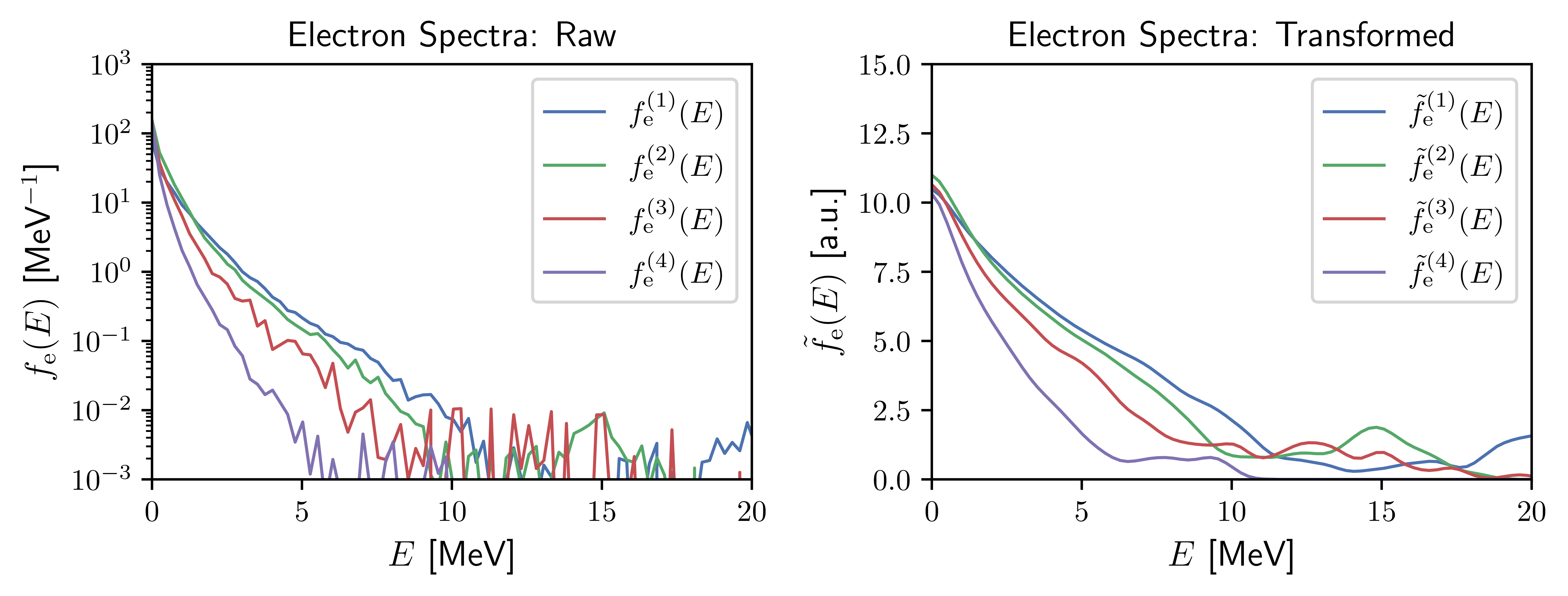}
\centering
\caption{Examples of electron spectra. Left: Raw spectra $f_\mathrm{e}(E)$ as measured in the PIC simulations (note the logarithmic scale). Right: Nonlinearly transformed spectra $\tilde{f}_\mathrm{e}(E)$.}
\label{fig:electron_spectra}
\end{figure}

In each simulation, we measured the electron spectra $500\si{fs}$ after the laser maximum reaches the target.
The electron spectra were computed by binning the weights of electron macroparticles onto $200$ bins with energies between 0 and $40\si{MeV}$.
Furthermore, aiming to bridge the more than five orders of magnitude present in the raw electron spectra, we made a nonlinear transformation $\tilde{f}_\mathrm{e}(E) = \left(\log{(f_\mathrm{e}/10^{-3}+1)} * g\right)(E)$,
which ensures that $\min_{E}{\tilde{f}_\mathrm{e}(E)} \geq 0$.
Here, $*$ relates to the convolution operation, i.e. we smoothed our spectra with a Gaussian filter $g(E)=\frac{1}{\sqrt{2 \pi}} e^{-E^2/2}$ to make the data more robust.
A comparison of the raw electron spectra to the transformed spectra is provided in Fig.~\ref{fig:electron_spectra}.

Given the limited data size, especially in relation to the relatively high number of bins, we experienced in our initial attempts of training a ML model from parameter vectors\footnote{Note that we use, for the sake of better of readability, henceforth the symbol $\vb{x}$ both for our simulation parameter $\vb{x}$ as well as normalized ML parameter $\hat{\vb{x}}$.} $\vb{x}$ to electron spectra $\tilde{f}_\mathrm{e}(E)$ that the training process is rather difficult and sensitive to hyperparameter.
Therefore, in order to further simplify the training process, we expressed our transformed spectra in terms of a linear regression model.
In particular, we performed principle component regression (PCR) using $n_{\vb{y}}=6$ principle components $b_k(E)$:
\begin{equation}
    \tilde{f}_\mathrm{e}(E) \approx \sum_{k=1}^{6}{c_k b_k(E)} + \bar{\tilde{f}}_\mathrm{e},
\end{equation}
where the $c_k$'s are the coefficients corresponding to the basis functions $b_k(E)$, and $\bar{\tilde{f}}_\mathrm{e}$ is the mean transformed spectrum.
Thus, altogether we train our ML models to learn mappings between $\vb{x}$ and $\vb{y} := \left[c_1, ..., c_6 \right]^\mathsf{T}$.
Further information about the ML models used and studied in this work and their training is provided in Sections~\ref{sec:inn} and~\ref{sec:results}, respectively.

\section{Invertible Neural Networks}\label{sec:inn}
In this work, we employ invertible neural networks (INNs) as described by Ardizzone et al. in~\cite{ardizzone2018analyzing}.
They have been designed having in mind a common problem in natural sciences, namely that there exists a forward process (experiment, simulation, ...) $f$ that maps a parameter vector $\vb{x} \in \mathbb{R}^{n_{\vb{x}}}$ onto experimentally accessible quantities which we call \textit{observables} $\vb{y}=f(\vb{x}) \in \mathbb{R}^{n_{\vb{y}}} $.
Typically, this forward process, at least formally, is well understood in the sense that there exists a (often highly sophisticated) theory that supports this mapping.
However, one is most often interested in the \textit{inverse process}, i.e. to gain information about experimentally inaccessible parameter given an experimental result.
Furthermore, since the forward process intrinsically has in general accompanying information loss, the inverse direction can only be expressed probabilistically as this loss of information renders the inverse process ambiguous.
In other words, multiple parameter vectors may correspond to the same observable.
Therefore, we are interested in the complete set of solutions $\{\vb{x} \in \mathbb{R}^{n_{\vb{x}}} \, \lvert \,  f(\vb{x}) = \vb{y}\}$, i.e. the full \textit{conditional posterior distribution} $p(\vb{x} \lvert \vb{y})$ has to be determined.

In order to account for the information loss, INNs introduce a \textit{latent space} of dimension $n_{\vb{z}}$ and elements $\vb{z} \sim \mathcal{N}(\vb{z}; \vb{0},\vb{I})$.
Then, the latent vectors are concatenated with the observables as $[\vb{y},\vb{z}]$.
Note that invertibility requires that $n_{\vb{x}} = n_{\vb{y}} + n_{\vb{z}}$, which can be realized by including zero-padding as required.
In our case, since already $n_{\vb{x}}=5 < n_{\vb{y}}=6$, we fulfilled the aforementioned condition by padding our parameter vectors as $[\vb{x}, \vb{0}]$, where $\vb{0}$ stands for a $(n_{\vb{y}} + n_{\vb{z}} - n_{\vb{x}})$-dimensional zero vector.
Then, INNs attain invertibility by composition of \textit{affine coupling blocks} (ACBs), which are invertible themselves.
While various different architectures for ACBs have been developed, in this work we use the Glow architecture\footnote{Not including ActNorm, invertible 1x1 convolutions, etc. relevant for their specific application, but only the coupling part itself.}, that is very similar on the RealNVP design~\cite{kingma2018glow,dinh2016density}.
In each ACB, the input is split into two parts $\vb{u}=[\vb{u}_1, \vb{u}_2]$ of equal size which are then transformed by an affine function using element-wise multiplication ($\odot$) and vector addition to an output $\vb{v}=[\vb{v}_1, \vb{v}_2]$:
\begin{equation}
\begin{split}
    \vb{v}_1 &= \vb{u}_1 \odot \exp(s_2(\vb{u}_2)) + t_2(\vb{u}_2), \\
    \vb{v}_2 &= \vb{u}_2 \odot \exp(s_1(\vb{u}_1)) + t_1(\vb{u}_1).
\end{split}
\end{equation}
Then, given the output $\vb{v}=[\vb{v}_1, \vb{v}_2]$, we can easily retrieve $\vb{u}=[\vb{u}_1, \vb{u}_2]$ as follows:
\begin{equation}
\begin{split}
    \vb{u}_2 &= (\vb{v}_2 - t_1(\vb{v}_1)) \odot \exp(-s_1(\vb{v}_1)), \\
    \vb{u}_1 &= (\vb{v}_1 - t_2(\vb{u}_2)) \odot \exp(-s_2(\vb{u}_2)).
\end{split}
\end{equation}
The functions $[s_i(\cdot), t_i(\cdot)]$, which are typically implemented as feedforward neural networks and hence called \textit{subnetworks}, can be arbitrarily complicated functions that need not be invertible themselves.
We further elaborate on the design of the subnetworks in Section~\ref{sec:results}.

INNs are bi-directionally trained with losses $\mathcal{L}_{\vb{x}}, \mathcal{L}_{\vb{y}}$ and $\mathcal{L}_{\vb{z}}$ defined for $\vb{x}, \vb{y}$ and $\vb{z}$, respectively.
While $\mathcal{L}_{\vb{y}}$ in general can be any supervised loss, we use the mean-squared loss (MSE) loss, $\mathcal{L}_{\vb{y}} = \mathbb{E}[(\vb{y} - f_{\vb{y}}(\vb{x}))^2]$.
For $\mathcal{L}_{\vb{x}}$ and $\mathcal{L}_{\vb{z}}$ we use \textit{maximum mean discrepancy} (MMD), which is a kernel-based, unsupervised loss on the space of probability distributions and which is based on reproducing kernel Hilbert spaces~\cite{gretton2012kernel}.
For our study, we used a multiscale inverse multiquadratic kernels as follows $k(\vb{x},\vb{x}') = \sum_{h}{1/(1 + \| (\vb{x} - \vb{x}')/h \|_2^2)}$, where the $\textit{bandwidth parameter}$ $h\in \{0.04,0.16,0.64\}$ are similar to the ones employed by Ardizzone et al.~\cite{ardizzone2018analyzing}.

\section{Results}\label{sec:results}
We ran our experiments on the Taurus cluster at ZIH/TU Dresden.
We used nodes of type Haswell, each node having two Intel Xeon E5-2680v3 @ 2.50 GHz
processors with 30 MB L3 cache and 12 cores each, amounting to 24 cores per node.
Each observation (1000 altogether), i.e. electron spectrum, was analyzed with one core.
Each core has $2 \cdot 32$KB L1 cache and 256 KB L2 cache.
Each program is written in Python 3.9.12 and imports NumPy 1.21.5 and PyTorch 1.10.2.

We performed two different experiments: \textbf{(1)} we made a comparison of acceptance rates between different methods for solving the inverse problem and \textbf{(2)} then measured the actual time needed to find one accepted solution.
These experiments were performed on hyperparameter optimized models as follows:

\subsubsection{ML Models and Training}
In this study, we both employ a multilayer perceptron (MLP) that we use as our reference model for the forward process $f(\cdot)$ only, and an INN for solving the inverse problem.
For the training of the MLP, we again use MSE loss, corresponding to the $\mathcal{L}_{\vb{y}}$ loss of the INN.
For both models we splitted our data into $80\%$ train and $20\%$ test set\footnote{I.e., 4000 and 1000 data points for the train and test set, respectively.}, and we used in both cases the \textit{Adam} optimizer with learning rate $\alpha=0.001$ and betas $\beta_1=0.9$, $\beta_2=0.999$ for training~\cite{kingma2014adam}.
Furthermore, we performed a hyperparameter optimization for the MLP with regards to:
\begin{enumerate}
    \item the activation function $\sigma(\cdot)$: ReLU($\cdot$), Tanh($\cdot$),
    \item widths of hidden layers: 12, 16, 20, 24, 30,
    \item number of layers: 3, 4, 5,
\end{enumerate}
where we found that the setting MLP: \{Tanh($\cdot$), 16, 4\} shows the lowest loss for the test set.
For the INN, we extend the hyperparameter optimization with regards to the dimension of the latent space $n_{\vb{z}}$, and the number of affine coupling blocks (ACBs):
\begin{enumerate}
    \item the activation function $\sigma(\cdot)$: ReLU($\cdot$), Tanh($\cdot$),
    \item widths of layers in subnetworks: 12, 16, 20, 24, 30,
    \item number of layers in subnetworks: 2, 3, 4,
    \item dimension $n_{\vb{z}}$: 2, 3, 4, 5, 6,
    \item number of ACBs: 2, 3, 4, 5, 6,
\end{enumerate}
where the best results, in terms of $\mathcal{L}_{\vb{x}} + \mathcal{L}_{\vb{y}}$, were obtained with the setting INN: \{Tanh($\cdot$), 20, 3, 6, 5\}.
Note that the optimal number of ACBs in our case is larger than proposed by Dinh who generally suggests to use four ACBs~\cite{dinh2014nice}.

\subsection{Acceptance Rate}
In the first experiment we compared, in terms of their acceptance rates, different methods suitable for (approximately) solving the inverse problem, i.e. to find a set of samples $\{\vb{x}\}$ that is representative for the conditional posterior $p(\vb{x} \lvert \vb{y}^\star)$ conditioned on a specific measurement $\vb{y}^\star$.
We call a parameter vector $\vb{x}$ to be accepted if the acceptance condition:
\begin{equation}
    d(\vb{y}^\star, f(\vb{x})) \leq \epsilon,
\end{equation}
is fulfilled, where $f(\cdot)$ is, in this study, the hyperparameter optimized MLP, $d(\cdot, \cdot)$ is a suitable distance function and $\epsilon$ is a non-negative threshold.

\begin{figure}[t]
\includegraphics[width=0.7\textwidth]{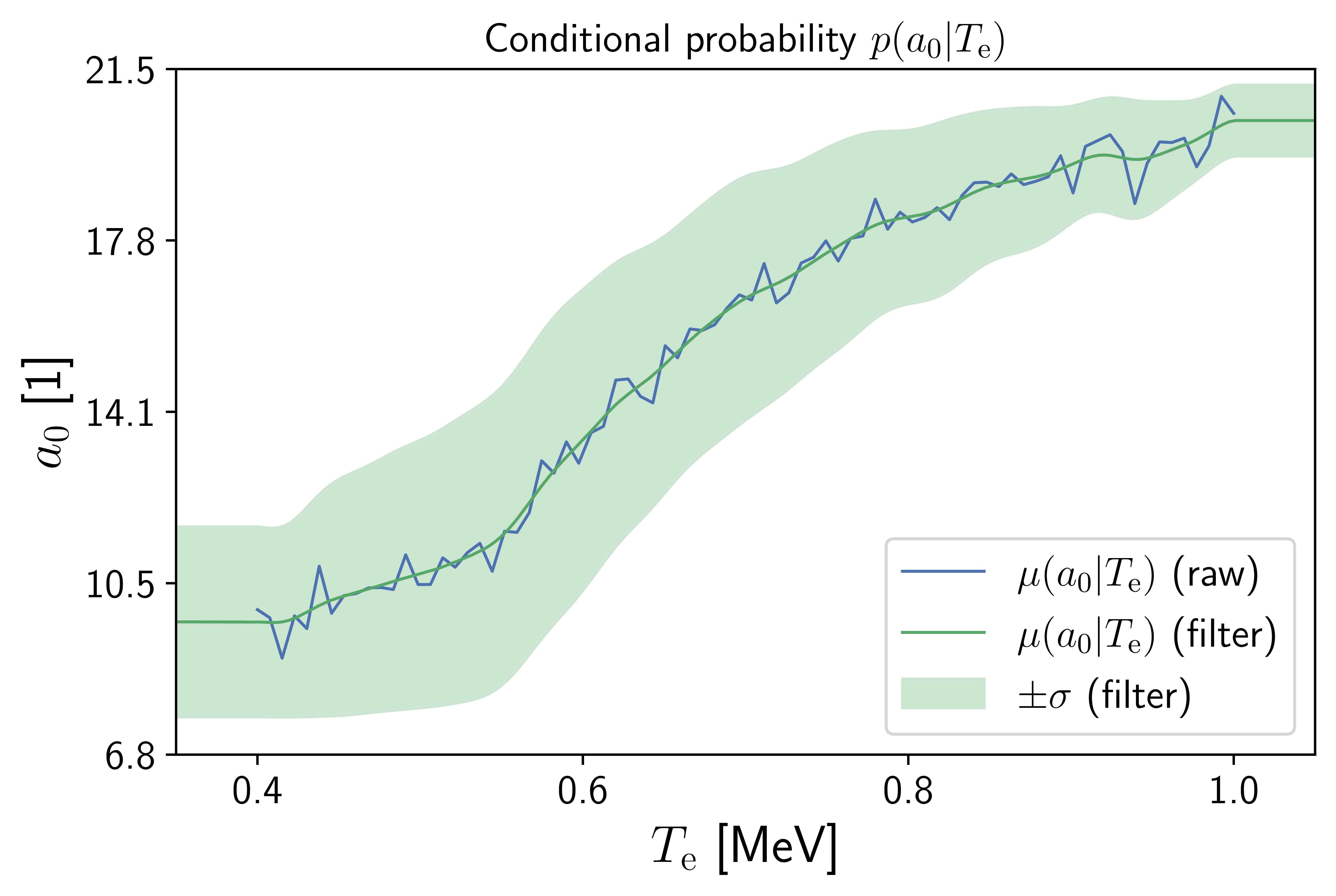}
\centering
\caption{Conditional probability $p(a_0 \lvert T_\mathrm{e})$ used as prior in the modified ABC routine.}
\label{fig:conditional_prob_a0_Te}
\end{figure}

\subsubsection{Approximate Bayesian Computation} If a surrogate model for the forward process is available, then one standard method to find an approximate solution for the inverse problem is \textit{approximate Bayesian computation} (ABC)~\cite{beaumont2019approximate}.
In ABC, the forward process $f$ is embedded in a rejection-sampling scheme, i.e. the forward model is employed by randomly sampling $\vb{x}$ from the parameter space and subsequently checking for the acceptance condition.

In practice, depending on the problem's complexity as well as $\vb{y}^\star$ and $\epsilon$, finding an appropriately sized set of solutions might require the evaluation of $f$ for millions of times.
Furthermore, since the algorithm is, per construction, subject to randomness, one can simply get "unlucky", consequently spending excessive amounts of compute time.
Therefore, in order to speed up the expected convergence, we also performed an experiment where we again used quasi-random numbers produced from the Halton sequence instead of purely randomly generated numbers.

Moreover, since in naive ABC we don't include prior knowledge as we draw samples (quasi)randomly from the parameter space and therefore implicitly assume a multivariate uniform distribution as our prior, we don't sample optimally and thus increase the computational effort.
Consequently, in order to study the effect of a non-uniform prior, we also conducted an experiment in which we draw $a_0$ based on a probability density function $p(a_0\lvert T_\mathrm{e})$, since $T_\mathrm{e}=T_\mathrm{e}(I(a_0))$. The corresponding probability distribution is illustrated in Fig.~\ref{fig:conditional_prob_a0_Te} and was numerically computed by applying Bayes' rule on the train set data.

\subsubsection{Hill-Climbing} On the other hand, instead of randomly trying different parameter vectors $\vb{x}$ as in ABC, local search algorithms such as hill-climbing (HC) and related methods try to find solutions by incrementally improving the current state~\cite{russel2013artificial}.
While typically gradient-based approaches are preferred, hill-climbing can be used also if only a black-box model is available.
Arguably it's simplest form, and also as implemented in this work, is first-choice hill-climbing, where the current solution is updated directly as soon as a better candidate solution has been found.
Considering that ABC produces a set of uncorrelated samples, however, it is necessary in HC, once having found the first solution, to restart the search for the next solution at a randomly chosen location in order also obtain a statistically uncorrelated sample.
A pseudocode of our implementation of first-choice hill-climbing is provided in the procedure FirstChoiceHillClimbing in Algorithm~\ref{alg:innhc}.
For our experiments, we used a learning rate of $\alpha=10^{-3}$, as commonly used in ML, and a learning rate of $\alpha=10^{-2}$ for comparison.

\subsubsection{Comparison}

\begin{figure}[t]
\includegraphics[width=\textwidth]{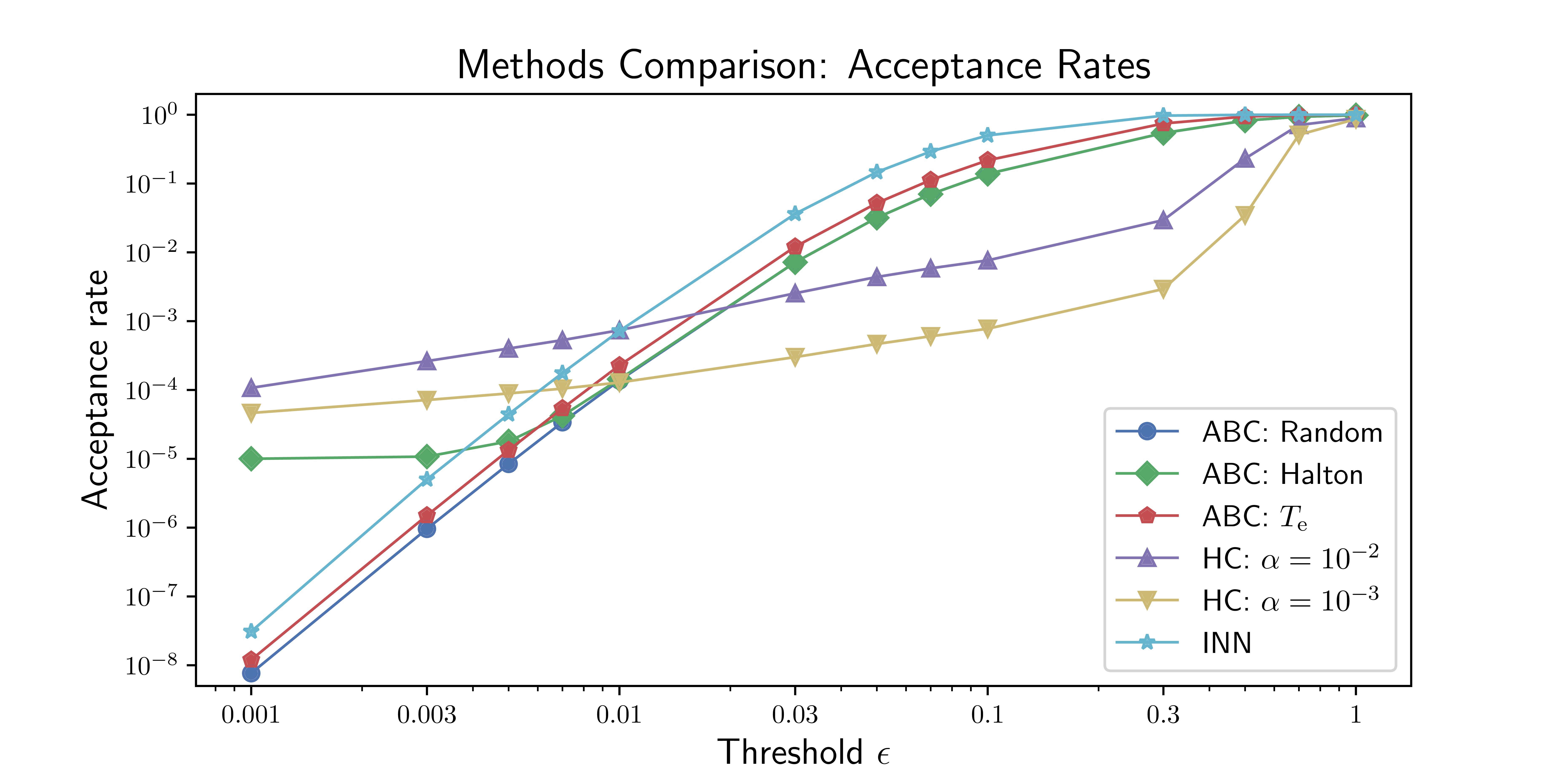}
\caption{Acceptance rates of different approaches for obtaining $p(\vb{x} \lvert \vb{y}^\star)$ in dependence of acceptance threshold $\epsilon$, averaged over 1000 different electron spectra.}
\label{fig:acceptance_rates}
\end{figure}

A comparison of the three different methods (ABC, HC and INN) and their specific settings is provided in Fig.~\ref{fig:acceptance_rates}.
First, we chose our test set of 1000 electron spectra $f_\mathrm{e}(E)$ and computed their PCR coefficients as described in Section~\ref{sec:sim_data}. Second, with the intention of obtaining a \textit{relative} measure of similarity, we define our distance function $d(\cdot, \cdot)$ based on the $L^2$ distance as follows:
\begin{equation}
    d[f(E), g(E)] = \frac{\sqrt{\int{(f(E) - g(E))^2 \dd{E}}}}{\sqrt{\int{g^2(E) \dd{E}}}},
\end{equation}
where $f(E)$ represents an electron spectrum containing errors, e.g. as proposed by one of our inverse solver, and $g(E)$ is the reference ground truth spectrum function. In this work, $g(E)$ corresponds to the transformed electron spectrum of the reference model, i.e. after the PCR procedure, and the lower and upper limits of integration are again $0\si{MeV}$ and $40\si{MeV}$, respectively.
At last, using 10 iterations with $m_\mathrm{trial}=10^5$, we compute the acceptance rate as the ratio of accepted solutions to all($=10\cdot m_\mathrm{trial}=10^6$) tested solutions.

From Fig.~\ref{fig:acceptance_rates} can see that the efficiency of a method heavily depends on the acceptance threshold $\epsilon$.
Not surprisingly, naive ABC using random numbers shows a rather low acceptance rate regardless of $\epsilon$.
Furthermore, ABC using quasi-random numbers behaves the same as naive ABC for larger $\epsilon$, since then the law of large numbers becomes relevant, and shows significantly better performance for smaller thresholds where $\epsilon \leq 5 \cdot 10^{-3}$.
A further improvement can be achieved by using an informed prior for $a_0$: On average, the acceptance rate increases approximately by a factor of 2 when compared to uninformed ABC.
However, interestingly, it can be seen that the acceptance rate is more than two orders of magnitude smaller than simply using ABC with the Halton sequence for $\epsilon = 10^{-3}$.
This can be understood by noting that in this case parameters are again sampled randomly, and thus don't exhibit the enhanced space-filling property as in the Halton case.
The largest acceptance rates for small $\epsilon$ were obtained by the hill-climbing methods, which are higher by around one order of magnitude.
We can deduce, from the big drop in the acceptance rate of HC for large $\epsilon$, that HC first needs many steps to approach a region of reasonably small distance after which, however, it apparently only takes minimal effort to further optimize the solution.
While HC with learning rate $\alpha=10^{-2}$ always beats HC with learning rate of $\alpha=10^{-3}$, we can also see that the difference significantly decreases for decreasing $\epsilon$.
On the other hand, the largest acceptance rates for $\epsilon \geq 10^{-2}$ were obtained by the INN.
For $\epsilon \leq 5\cdot 10^{-3}$, the INN shows worse performance than both configurations of HC.
Then, around $\epsilon \approx 3\cdot 10^{-3}$, the INN also exhibits lower acceptance rates than Halton-based ABC.
Again, the reason is that latent vectors $\vb{z}$ are sampled randomly from the multivariate normal distribution, and not from a quasi-random sequence.
Therefore, for the full range of $\epsilon$, the INN always surpasses the acceptance rate of ABC Random and ABC $T_\mathrm{e}$, since they are also both based on random numbers rather than a quasi-random sequence.

\subsection{Runtimes}
\begin{figure}[t]
\includegraphics[width=\textwidth]{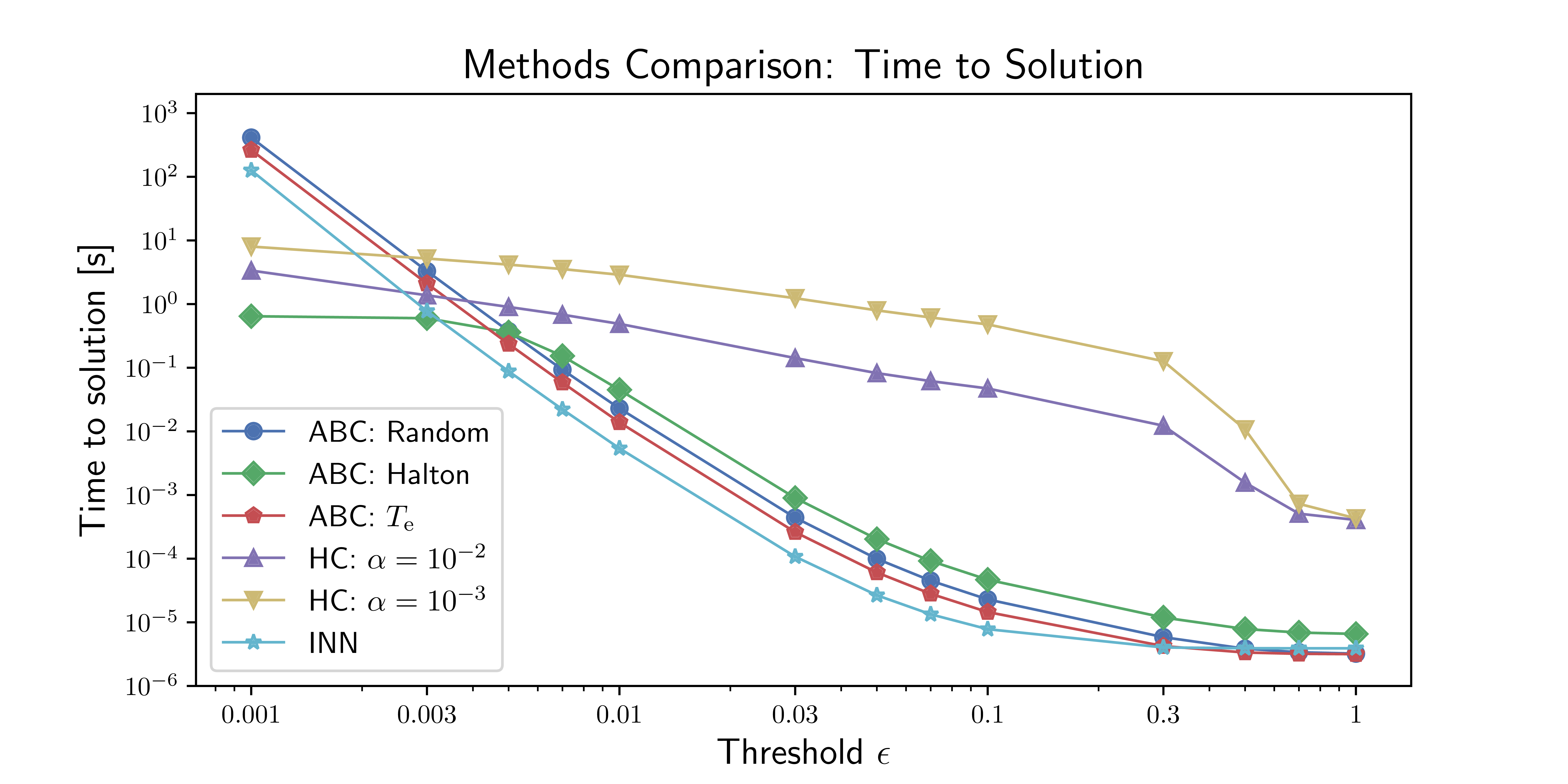}
\caption{Time to solution, i.e. of a single acceptance, of different approaches for obtaining $p(\vb{x} \lvert \vb{y}^\star)$ in dependence of acceptance threshold $\epsilon$, averaged over 1000 different electron spectra.}
\label{fig:time_sol}
\end{figure}

While the acceptance rate is more interesting for theoretical analysis, in practice we are interested in the actual computational cost, e.g. in terms of the runtime.
We performed the same experiment as before, but instead we measured the total runtime relative to the number of accepted samples, $t_\mathrm{total} / m_\mathrm{acc}$.
Again, we average our results over the same 1000 electron spectra from the test set.
The result is depicted in Fig~\ref{fig:time_sol}.
We can see that general trends are preserved, e.g. that the INN is the fastest method up until $\epsilon \geq 3\cdot 10^{-3}$ and that ABC with informed prior is always faster than naive ABC.
HC is still a fast method for very small thresholds.
However, the relative runtimes difference between HC and all other methods decreased by around two order of magnitude when compared to the relative difference in acceptance rate.
This can be understood as HC uses loops and needs to call $f(\cdot)$ many times, while e.g. ABC calls the forward function only once for all proposal vectors and therefore takes advantage of optimized matrix operations.
Hence, to ensure efficient computation, HC should as well be embedded in parallel procedures.
We can also see that using the INN does not cause any significant overhead, since the relative speedup is approximately conserved with respect to the relative increase in acceptance rate.

Thus, in order find uncorrelated samples for the inverse problem, the method should be chosen having the required accuracy in mind.
An algorithm based on the combination of an INN and HC, which is designed to also work for very small thresholds, is provided in Algorithm~\ref{alg:innhc}.

\begin{algorithm}[t]
    \caption{INN-HC: Inverse solver optimized for low acceptance thresholds.}\label{inverse}
    \hspace*{\algorithmicindent}
    \begin{algorithmic}[1]
    \Procedure{InverseSolver}{$\vb{y}^\star$, $m$, $f$, $\mathrm{inn}$, $d$, $\epsilon$, $\alpha$}
    \State $\vb{Y}^\star \gets \mathrm{vstack}(\vb{y}^\star, m)$\Comment{Vertically stack $\vb{y}^\star$, i.e. $\vb{Y}^\star \in \mathbb{R}^{m \cross n_{\vb{y}}}$}
    \State $\vb{Z} \gets \mathrm{rand}(\mathcal{N}(0,1), (m, n_{\vb{z}}))$\Comment{$\vb{Z} \in \mathbb{R}^{m \cross n_{\vb{z}}}$}
    \State $\vb{X} \gets \mathrm{inn}^{-1}([\vb{Y}^\star, \vb{Z}])$
    \For{$i$ in $1,...,m$}
    \State $\vb{x} \gets \vb{X}_{i,\boldsymbol{\cdot}}$
    \If {$d(\vb{y}^\star, f(\vb{x})) > \epsilon$}
    \State $\vb{x} \gets \mathrm{FirstChoiceHillClimbing}(\vb{y}^\star,f,d,\epsilon,\vb{x},\alpha)$
    \State $\vb{X}_{i,\boldsymbol{\cdot}} \gets \vb{x}$
    \EndIf
    \EndFor
    \State \textbf{return} $\vb{X}$
    \EndProcedure
    \Statex
    \Procedure{FirstChoiceHillClimbing}{$\vb{y}^\star$, $f$, $d$, $\epsilon$, $\vb{x}_0$, $\alpha$}
    \State $\vb{x} \gets \vb{x}_0$
    \While{$d(\vb{y}^\star, f(\vb{x})) > \epsilon$}
    \State $\bm{\xi} \gets \mathrm{rand}(\mathcal{U}([-1, 1]), (n_{\vb{x}})) $\Comment{Generate vector with random direction}
    \State $\bm{\xi} \gets \bm{\xi}/\lvert \bm{\xi} \rvert $\Comment{Normalize to unit length}
    \State $\tilde{\vb{x}} \gets \vb{x} + \alpha \bm{\xi}$
    \If {$d(\vb{y}^\star, f(\vb{\tilde{x}})) \leq d(\vb{y}^\star, f(\vb{x}))$}
    \State $\vb{x} \gets \tilde{\vb{x}}$.
    \EndIf
    \EndWhile
    \State \textbf{return} $\vb{x}$
    \EndProcedure
    \end{algorithmic}
\label{alg:innhc}
\end{algorithm}

\section{Conclusion}
In this work, we have studied INNs on synthetic electron spectra in the context of near-critical laser-plasma physics.
In particular, we compared INNs with other standard statistical methods for solving the inverse process.
We found that INNs perform, both in terms of acceptance rates as well as runtimes, better than all other methods up to a small threshold distance.
Furthermore, we show that naive ABC based on random numbers has lower acceptance rates and larger runtimes than our INN model by a factor of approximately 10 for any threshold.
INNs also surpass informed ABC, where we used a modified prior $p(a_0 \lvert T_\mathrm{e})$ that we motivated due to physical considerations, by a significant amount.
Moreover, we demonstrate the importance of quasi-random numbers and recommend to use them as well in conjunction with INNs.
On the other hand, our results suggest that iterative approaches, in our case hill-climbing, surpass INNs for small thresholds $\epsilon \leq 3 \cdot 10^{-3}$, especially in terms of the acceptance rate.
Therefore, due to the relative strengths of the different algorithms, we propose a composite algorithm for obtaining the conditional posterior that combines both hill-climbing and INNs.

\bibliography{bibliography}

\end{document}